\documentclass[a4paper]{article}
\usepackage[latin1]{inputenc}
\usepackage[T1]{fontenc}
\usepackage[dvips]{graphicx}
\begin{document}
\title{Fermat's principle and variational analysis of an optical model
for  light propagation exhibiting a critical radius}
\author{ M.\ Marklund, D.\ Anderson, F.\ Cattani, M.\ Lisak and L.\ Lundgren  \\[2mm]
  \textit{Department of Electromagnetics,
    Chalmers University of Technology,} \\
    \textit{SE--412~  96 Göteborg, Sweden}}
\date{(\today)}
\maketitle

\begin{abstract}
Fermat's principle and variational analysis is used to analyze the
trajectories of light propagating in a radially inhomogeneous
medium with a singularity in the center. It is found that the
light trajectories are similar to those around a black hole, in
the sense that there exists a critical radius within which the
light cannot escape, but  spirals into the singularity.
\end{abstract}

\section{Introduction}

There has recently, within the scientific community, been much
interest focused on the
extra-ordinary properties associated with Bose-Einstein
condensates - clouds of atoms cooled down to nano-Kelvin
temperatures where all atoms are in the same quantum state and
macroscopic quantum conditions prevail
\cite{Leonhardt-Piwnicki-99,Leonhardt-Piwnicki-00,Hau-etal-99}.

One of the most dramatic experiments which has been performed in
Bose-Einstein condensates is the demonstration of optical light
pulses traveling at extremely small group velocities, e.g.
velocities as small as 17 m/s have been reported
\cite{Hau-etal-99}.

In a physically suggestive application of the optical properties
of Bose-Einstein condensates, it was recently suggested that they
could be used to create, in the laboratory, dielectric analogues
of relativistic astronomical phenomena like those associated with
a black hole \cite{Leonhardt-Piwnicki-00}.

The astrophysical concept of a black hole is one of the most
fascinating and intriguing phenomena related to the interaction
between light and matter through the curvature of space-time
within the framework of general relativity. Black holes are
believed to form when compact stars undergo complete gravitational
collapse due to, e.g., accretion of matter from its surroundings
(in general relativity, the pressure contributes to the
gravitational mass of a fluid, and an increased pressure will,
after a certain point, therefore only help to accelerate the
collapse phase).

In ordinary dielectric media, unrealistic physical conditions
would be required in order to demonstrate any of the spectacular
effects of general relativity. However, in dielectric media
characterized by very small light velocities, new possibilities
appear. In particular, it was recently suggested that by creating
a
vortex structure in such a dielectric medium it is possible to
mimic the properties of an optical black hole
\cite{Leonhardt-Piwnicki-00}.
What this study in fact suggested was that one could construct,
by using the aforementioned vortex,  an unstable photon
orbit, much like the orbit at a radius $r = 3M$ in the
Schwarschild
geometry \cite{Misner-Thorne-Wheeler}. In order to construct an
event horizon, one would need to supplement the vortex flow by a
radial motion of the fluid
\cite{Visser-00,Leonhardt-Piwnicki-00b}.

The vortex, which involves a rotating cylindrical velocity field,
tends to attract light propagating perpendicular to its axis of
rotation and to make it deviate from its straight path. In the
model considered in Refs.\
\cite{Leonhardt-Piwnicki-99,Leonhardt-Piwnicki-00},
the increasing, in fact diverging,
rotational velocity field of the vortex core attracts light by the
optical Aharanov--Bohm effect and causes a bending of the light
ray
similar to that of the gravitational field in general relativity.
It is also shown that there exist a critical radius, $r_{crit}$,
from the vortex core with the properties that if light rays come
closer to the core than $r_{crit}$,the light will fall
towards the singularity of the vortex core. In this sense, the
critical radius, $r_{crit}$, plays the role of an optical
unstable photon orbit, analogous to the unstable orbit in the
Schwarzschild geometry.

The basic physical effect involved in the  analysis of the ray
propagation in Refs.\
\cite{Leonhardt-Piwnicki-99,Leonhardt-Piwnicki-00}
 is the fact that the refractive index, $ n$,
of a medium changes when the medium is moving. In Refs.\
\cite{Leonhardt-Piwnicki-99,Leonhardt-Piwnicki-00},
the medium is assumed to have a cylindrical vortex velocity
field $ \bf V= V(\bf r)$ given by
\begin{equation}
{\bf V(r)} = \frac{W}{r}{\bf \widehat{\varphi}}
\end{equation}
where $r$ is the radius from the vortex core centre,
$\widehat{\varphi }$ is the azimuthal unit vector and $2\pi W$ is
the vorticity.

Advanced physical concepts like Bose-Einstein condensates, vortex
velocity fields and black holes are fascinating physical effects,
which however may be difficult to present in a simplified manner
for  undergraduate students. Nevertheless, it is important to try
to convey inspiration for and arouse curiosity in such phenomena.
An interesting example of such an effort was recently made in
Ref.\ \cite{McDonald-00}, where the basic physical mechanism, in
the form of a classical two level system was used to demonstrate
the possibility of the very low light velocities observed in Bose-
Einstein condensates. In a later work by the same author,
 Ref. \ \cite{McDonald-01}, an inspiring  investigation is given of a
simple mechanical model that exhibits a gravitational critical
radius. The purpose of the present work is similar to that of
Refs.\ \cite{McDonald-00,McDonald-01}. We will try to discuss a
simple classical example of a medium where the refractive index
has a divergence at $r = 0$ and which exhibits some of the
dramatic properties of the light behaviour around a black hole as
discussed in Refs.\
\cite{Leonhardt-Piwnicki-99,Leonhardt-Piwnicki-00,Visser-00,%
Leonhardt-Piwnicki-00b}. In addition we
want to illustrate the power and beauty of  the classical
principle of  Fermat  and of variational methods in connection
with this new fascinating concepts.

\section{Fermat's principle}\label{sec:Fermat}

 It is well known that light
tends to be deflected towards regions with higher refractive
index. Consider light propagating in a cylindrically symmetric
medium where the refractive index increases towards the centre. In
such a situation we expect the light path to look qualitatively as
in Fig.\ 1. The actual light path is determined by Fermat's
principle i.e.\
\begin{equation}
  \delta\int n({\bf r})ds=0
\end{equation}
where$ ds$ is an infinitesimal element along the light ray. We
will consider light propagation in a medium where the refractive
index is of the qualitative form, cf Eq.\ (1)
\begin{equation}
  n(r)\simeq \left\{
  \begin{array} {r@{\quad as\quad}l}1 & r\gg r_{0}\\
  r_0/r & r\ll r_{0}\end{array} \right.
\end{equation}

One possible realization of $n(r)$ with the desired properties is
given by
\begin{equation}
  n^{2}(r) = 1 + \left(\frac{r_{0}}{r}\right)^2
\end{equation}

Using Fermat's principle and expressing the light path as the
relation $\theta=\theta(r)$ where $\theta$ is the polar angle,
Eq.\ (2) implies
\begin{equation}
  \delta\int{n(r)\sqrt{1+r^{2}\left(\frac{d\theta}{dr}\right)^2}}dr
  = 0 \ .
\end{equation}

The Euler-Lagrange variational equation corresponding to Eq.\ (5)
reads
\begin{equation}
  \frac{d}{dr}\left[ n(r)\frac{r^2 \, d\theta/dr}%
  {\sqrt{1 + r^2(d\theta/dr)^2}}\right]=0
\end{equation}
which determines the trajectory of the light.

\section{Solution of the light trajectory}

Equation (6) directly implies that
\begin{equation}
  \frac{r^2n(r) \, d\theta/dr }%
  {\sqrt{1 + r^2 (d\theta/dr)^2}} = r_{i}
\end{equation}
where the constant $r_{i}$ is determined by initial conditions,
i.e.\ the properties of the incident light.

Equation (7) is easily inverted to read
\begin{equation}
  \frac{d\theta}{dr}=\pm\frac{r_{i}}{\sqrt{r^4n^2(r) - r_{i}^2r^2}}
\end{equation}

For a light ray incident as shown in Fig.\ 1, we clearly have
$d\theta/dr <0$ (at least up to some minimum radius $ r=r_{min}$.
It is illustrative to first consider the trivial case of a
homogeneous medium with $n(r)\equiv1$. In this case Eq.\ (9)
becomes
\begin{equation}
  \frac{d\theta}{dr}=-\frac{r_{i}}{\sqrt{r^4 - r_{i}^2r^2}}
\end{equation}
which can easily be interpreted to yield the light path in the
form
\begin{equation}
  \theta=\mathrm{arccot}\sqrt{\frac{r^2}{r_{i}^2} - 1}
\end{equation}
or simpler
\begin{equation}
  r=\frac{r_{i}}{\sin \theta}
\end{equation}
Clearly this is the straight line solution
\begin{equation}
  y=r_{i}
\end{equation}
where the parameter $r_{i}$ plays the role of ``impact parameter''
or minimum distance from the centre.

Let us now consider the model variation for $n(r)$ as given by
Eq.\ (5). Within this model Eq.\ (10) becomes
\begin{equation}
  \frac{d\theta}{dr} =
  - \frac{r_i}{\sqrt{r^4 - (r_i^2 - r_0^2)r^2}}
\end{equation}
Clearly the solution of Eq.\ (13) will depend crucially on the
relative magnitude of $r_i$ and $r_0$, i.e. the impact parameter
relative to the characteristic radial extension of the
inhomogeneity. Consider first the case when $r_0 < r_i$. Equation
(13) can then be rewritten as
\begin{eqnarray}
  \frac{b_1}{r_i}\frac{d\theta}{dr}
  &=& -\frac{b_1}{\sqrt{r^4 -b_1^2r^2}} \ , \\
  b_1 &\equiv& \sqrt{r_{i}^{2}-r_{0}^{2}} \ .
\end{eqnarray}
Equation (15) is of the same form as Eq.\ (9) and we directly
infer the
following solution
\begin{equation}
  r = \frac{\sqrt{r_{i}^{2} - r_{0}^{2}}}%
  {\sin\left[\theta\sqrt{1-r_{0}^{2}/r_{i}^{2}}\,\right]}
\end{equation}

As $\theta\rightarrow0$, we still have asymptotically
\begin{equation}
  r\simeq\frac{r_{i}}{\sin \theta}
\end{equation}
However, the trajectory is now bending towards the origin and the
minimum distance occurs at the polar angle $\theta=\theta_{m}$
given by
\begin{equation}
  \theta_m = \frac{\pi}{2}\frac{1}{\sqrt{1-r_0^2/r_i^2}}
\end{equation}

The corresponding minimum distance, $r_m$ is
\begin{equation}
  r_m \equiv r{\theta_m} = \sqrt{r_i^2 - r_0^2}
\end{equation}

We also note that the trajectory is symmetrical around the angle
$\theta_{m}$ and that the asymptotic angle of the outgoing light
ray is
\begin{equation}
  \theta_{\infty}\equiv \lim_{r\rightarrow\infty} \theta(r)
  =\frac{\pi}{\sqrt{1 - r_0^2/r_i^2}} = 2 \theta_m
\end{equation}

The solution given by eq.(10) describes a trajectory which is bent
towards the centre of attraction at $ r=0$. Depending on the ratio
$r_{0}/r_{i}$, the trajectory is either more or less bent or may
even perform a number of spirals towards the centre before again
turning outwards and escaping, cf. Fig.\ 2. The number of turns,
$N$,which the light ray does around the origin before escaping is
simply
\begin{equation}
  N = \left\lfloor \frac{2\theta_{m}}{2\pi} \right\rfloor
    = \left\lfloor \frac{1}{2\sqrt{1-r_{0}^{2}/r_{i}^{2}}} \right\rfloor
\end{equation}
where $\left\lfloor x \right\rfloor$ denotes the largest integer
less than $x$.

Let us now consider the special case when the impact parameter
equals the characteristic width of the refractive index core i.e.\
$r_{i}=r_{0}$. The equation for the trajectory now simplifies to
\begin{equation}
  \frac{d\theta}{dr} = -\frac{r_{i}}{r^{2}}
\end{equation}
with the simple solution
\begin{equation}
  \theta=\frac{r_{i}}{r}
\end{equation}
i.e.\ the trajectory describes a path in the form of Arkimede's
spiral as the light falls towards the origin, cf. Fig.\ 3. The
form of the light trajectory in the situation when $ r_{i}< r_{0}$
is now obvious, it will spiral into the singularity more or less
directly depending on the magnitude of the ratio $r_{0}/r_{i}>1$.

The actual trajectory in this case is given by a slight
generalization of Eq.\ (16) viz
\begin{equation}
  r=\frac{\sqrt{r_{0}^{2}-r_{i}^{2}}}{\sinh
  [\theta\sqrt{1-r_{i}^{2}/r_{0}^{2}}]}
\end{equation}

This solution does indeed convey the expected behaviour, the
trajectory spirals monotonously into the  singularity of the
refractive index.

\section{Final comments}

The present analysis is inspired by recent discoveries and
discussions about light propagation in Bose-Einstein condensates
where extremely low light velocities can be obtained.
This has triggered speculations about possible laboratory
demonstrations
of effects, which normally are associated with general
relativistic
conditions. In particular, it has been suggested that it
may be possible to create the analogue of a black hole using a
divergent in-spiral of a Bose-Einstein condensate.

In the present work we have analyzed a simple classical example of
light propagation as determined by Fermat's principle in a medium
characterized by a radially symmetric refractive index. In analogy
with the variation of the vortex velocity field suggested in
\cite{Leonhardt-Piwnicki-00}, the refractive
index is here assumed to diverge towards the centre. This
classical
example exhibit some of the characteristic properties of light
propagating around a black hole where the gravitational attraction
deflects the light and where, under certain conditions, the light
may be "swallowed" by the black hole.

In the example analyzed here, the unstable photon orbit of
the Schwarzschild black hole is similar to the characteristic
radius,
$r_0$, of the refractive index variation, which together with the
impact parameter, $r_i$, of the incident light completely
determines the light trajectory. If $r_0 < r_i$, the light is more
or less deflected, but ultimately escapes. However, if
$r_0 \geq r_i$, the light spirals into the singularity.\
It should be cautioned that this result depends crucially on the
presence of the singularity in the refractive index (4). If this
is removed the in-spiraling photon orbit will eventually turn and
start spiraling outwards.  In the recent discussion about the
possibility of generating analogues of optical black holes in
Bose-Einstein condensates, it has been suggested that the vortex
motion must also have a velocity component in the radial
direction.

In the classical model considered here,  a number of additional
physical effects will obviously affect the light path when it
comes close to the axis and will in fact remove the mathematical
singularity. Nevertheless, the model provides a simple example of
light dynamics, which resembles some of the properties of light
propagation around a black hole. Furthermore,the investigation is
based on Fermat's principle and variational analysis, in this way
illustrating the use of classical methods in connection with very
new and fascinating concepts at the front line of modern research.

 \section*{Figure Captions.}
 Fig.1   Qualitative plot of a light ray trajectory in a cylindrically symmetric medium with a refractive
 index, which increases  towards the centre.\\
 Fig.2  Light trajectories for the refractive index model of eq.()
for different impact radii $r_{i}>r_{0}$\\ Fig.3  Light trajectory
in the case of $r_{i}=r_{0}$, the spiral of Arkimede.\\


\begin{thebibliography}{99}

  \bibitem{Leonhardt-Piwnicki-99}
    U.\ Leonhardt and P.\ Piwnicki,
    Phys.\ Rev.\ A \textbf{60}, 4301 (1999)

  \bibitem{Leonhardt-Piwnicki-00}
    U.\ Leonhardt and P.\ Piwnicki,
    Phys.\ Rev.\ Letters \textbf{84}, 822 (2000)

  \bibitem{Hau-etal-99}
    L.\ V.\ Hau, S.\ E.\ Harris, Z.\ Dutton and C.\ H.\ Behroozi,
    Nature (London) \textbf{397}, 594 (1999)

  \bibitem{Misner-Thorne-Wheeler}
    C.\ Misner, K.\ S.\ Thorne and J.\ A.\ Wheeler \textit{Gravitation}
 (Freeman, 1973)
  \bibitem{Visser-00}
    M.\ Visser,
    Phys.\ Rev.\ Letters \textbf {85}, 5252 (2000)

  \bibitem{Leonhardt-Piwnicki-00b}
    U.\ Leonhardt and P.\ Piwnicki,
   Phys.\ Rev.\ Letters \textbf {85}, 5253 (2000)


  \bibitem{McDonald-00}
    K.\ T.\ Mc Donald, Am.\ J.\ Phys.\ \textbf{68}, 293 (2000)
\bibitem{McDonald-01}
    K.\ T.\ Mc Donald, Am.\ J.\ Phys.\ \textbf{69}, 617 (2001)
\end{thebibliography}
\end{document}